\documentclass[12pt]{article}
\begin{document}
\title{Noncommutative Geometry and Some Issues}
\author{B.G. Sidharth\\
International Institute for Applicable Mathematics \& Information Sciences\\
B.M. Birla Science Centre, Adarsh Nagar, Hyderabad - 500 063 (India)}
\date{}
\maketitle
\begin{abstract}
Recent observations of Ultra High Energy Cosmic rays suggest a small violation of Lorentz symmetry. Such a violation is expected in schemes with discrete/quantized spacetime. We examine this situation and suggest tests which could be carried out, for example by NASA's GLAST Satellite. The considerations are extrapolated to the large scale cosmos.
\end{abstract}
\section{Introduction}
A noncommutative geometry arises if there is a minimum spacetime interval that cannot be penetrated, as is well known, starting from the work of Snyder and others \cite{snyder1,sny}. Later interest in such minimum intervals was aroused due to, for example, lattice gauge theory \cite{Wilson}, though it was a computational tool. This apart these ideas were also studied independently by several authors like Kadyshevskii, Wolf, Finkelstein, the author himself and others, though in the context of differing settings \cite{kady,wolf,fink,cu}. Even more recently 't Hooft and others have re-examined lattice theories. This time the motivation has been more on the lines of minimum spacetime intervals \cite{tooft}.\\
What happens in this case is there is a departure from Lorentz Symmetry \cite{cu,mont}. Typically we have an energy momentum relation (with units such that $c=1=\hbar)$
\begin{equation}
E^2 = m^2 + p^2 - l^2 p^4\label{e1}
\end{equation}
where $l$ is a minimum length interval, which could be typically the Planck length and more generally the Compton length, (which reduces to the Planck length for a Planck mass). Interestingly we could arrive at (\ref{e1}) from an alternative point of view, starting directly from the noncommutativity, or the modified uncertainty principle which results from these considerations in Quantum Gravity or Quantum Superstring theory, for example (cf.\cite{mup}and references therein):
\begin{equation}
[x,p] = \hbar' = \hbar [1 + \left(\frac{l}{\hbar}\right)^2 p^2]\, etc\label{e2}
\end{equation}
where we have temporarily re-introduced $\hbar$. (\ref{e2}) shows that effectively $\hbar$ is replaced by $\hbar'$. So,
$$E = (m^2 + p^2)^{\frac{1}{2}} (1 + l^2 p^2)^{-1}$$
or
\begin{equation}
E^2 = m^2 + p^2 - 2l^2 p^2,\label{e3}
\end{equation}
neglecting higher order terms in (\ref{e3}) is of the same form as (\ref{e1}).\\
We now examine a few implications of (\ref{e1}).
\section{Modified Dispersion}
Let us consider an effect similar to the Compton effect \cite{powell}, but with (\ref{e1}) replacing the usual energy momentum formula. Here if $\vec k_0$ is the incident radiation and $\vec k$ is the scattered radiation at an angle $\theta$, as in the usual theory we get from the energy and momentum conservation laws,
\begin{equation}
k_0 - k = E-m\label{e4}
\end{equation}
and
\begin{equation}
\vec k_0 - \vec k = \vec p\label{e5}
\end{equation}
Further algebraic manipulation of (\ref{e4}) and (\ref{e5}) gives
$$kk_0 (1-\cos \Theta ) = m (k_0-k) + \frac{l^2}{2} [Q^2 + 2mQ ]^2 = mQ + \frac{l^2}{2}[Q^2 + 2mQ]^2$$
where
$$E-m = Q = k_0-k$$
Whence, we get the frequency $k$ as,
\begin{equation}
k = \frac{mk_0 + \frac{l^2}{2} [Q^2 + 2mQ]^2}{[m+k_0 (1-\cos \Theta )]}\label{e6}
\end{equation}
Alternatively, let us denote the additional change in frequency due to the noncommutativity of spacetime or the presence of minimum spacetime intervals by $\epsilon$, so that
$$k + \epsilon = \bar k$$
$\bar k$ being the usual compton frequency. With this, we get, instead of (\ref{e6}),
\begin{equation}
\epsilon  = \frac{l^2 [Q^2 + 2mQ]^2}{2\{m+k_0(1-\cos \Theta )\}}\label{e7}
\end{equation}
The relations (\ref{e6}) or (\ref{e7}) enable us to observe the effect of the violation of Lorentz Symmetry as embodied in (\ref{e1}).
\section{Particle Behaviour}
Owing to (\ref{e1}) we have a modified Klein-Gordon equation
\begin{equation}
(D + l^2 \nabla^4 - m^2)\psi = 0\label{e8}
\end{equation}
where $D$ denotes the usual D'Alembertian.\\
Just to get a feel, it would be interesting to consider the extra effect in (\ref{e8}). For simplicity we take the one dimensional case. As in conventional theory if we separate the space and time parts of the wave function we get
\begin{equation}
l^2 u^{(4)} + u^{(2)} + \lambda u = 0, \quad \lambda = E^2 - m^2 > 0\label{e9}
\end{equation}
whence if in (\ref{e9}) we take,
$$u = e^{\alpha x}$$
and  $\alpha^2 = \beta$ we get,
$$l^2 \beta^2 + \beta + \lambda = 0$$
whence
$$\beta = \frac{-1 \pm \sqrt{1-4l^2\lambda}}{2l^2}$$
So
\begin{equation}
\beta \approx \frac{-1 \pm \{1 - 2l^2 \lambda\}}{2l^2}\label{e10}
\end{equation}
From (\ref{e10}) it is easy to deduce that there are two extra solutions, as can be anticipated by the fact taht (\ref{e8}) is a fourth order equation, unlike the usual second order Klein-Gordon equation. Thus we have
$$\beta = -\lambda (<0)$$
giving the usual solutions, but additionally we have
\begin{equation}
\beta = -\left(\frac{1 - \lambda l^2}{l^2}\right)(<0)\label{e11}
\end{equation}
What do the two extra solutions in (\ref{e11}) indicate? To see this we observe that $\alpha$ is given by, from (\ref{e11})
\begin{equation}
|\alpha | \approx \pm \frac{1}{l}\label{e12}
\end{equation}
In other words (\ref{e12}) corresponds to waves with wavelength of the order $l$, which is intuitively quite reasonable.\\
What is interesting is that if $l$ is an absolute length then the extra effect is independent of the mass of the particle. In any case the solutions from (\ref{e12}) are GZK violating solutions.\\
We now make some remarks. Departures from Lorentz symmetry of the type given in (\ref{e1}) have been studied, though from a phenomenological point of view \cite{mestres,glashow,jacob,olinto,carroll,nag}. These arise mostly from an observation of ultra high energy cosmic rays. Given Lorentz symmetry, there is the GZK cut off such that particles above this cut off would not be able to travel cosmological distances and reach the earth. However there are indications of a violation of the GZK cut off.\\
In any case some of the effects following, for example from (\ref{e1}), like (\ref{e6}) or (\ref{e7}) can be detected, it is hoped by the GLAST Satellite to be launched by NASA in 2006 or shortly thereafter.\\
Interestingly, if in (\ref{e1}) or (\ref{e8}) we take, purely on an ad hoc basis, $-l^2$ rather than $+l^2$, we get two real exponential solution of (\ref{e8}). One of them is an increasing exponential leading to very high probabilities for finding these particles.
\section{Scaled Fractality}
We now extrapolate the minimum scale effects to the large scale universe. It has been known that there is a deep connection between a stochastic and Brownian behaviour on the one hand and critical point phenomena and the Renormalization Group on the other hand \cite{Nottale1,Davis}. Fractality itself is a manifestation of resolution dependent measurements, while Renormalization Group considerations arise due to coarse graining at different resolutions. A good example of the fractal behaviour is Quantum Mechanics itself which has been shown to have the fractal dimension 2 \cite{Nottale2}.\\
In the above context, we will now argue that there is a manifestation of what may be called ``scaled'' Quantum Mechanics, at different scales in the universe.\\
It has already been argued that in the universe at large, there appears to be the analogues of the Planck constant at different scales \cite{BGSSU1,BGSSU2}. Infact we have
\begin{equation}
h_1 \sim 10^{93}\label{e13}
\end{equation}
for super clusters;
\begin{equation}
h_2 \sim 10^{74}\label{e14}
\end{equation}
for galaxies and
\begin{equation}
h_3 \sim 10^{54}\label{e15}
\end{equation}
for stars. And
\begin{equation}
h_4 \sim 10^{34}\label{e16}
\end{equation}
for Kuiper Belt objects. In equations (\ref{e13}) - (\ref{e16}),
the $h_\imath$ play the role of the Planck constant, in a sense to
be described below. The origin of these equations is related to
the following empirical relations
\begin{equation}
R \approx l_1 \sqrt{N_1}\label{e17}
\end{equation}
\begin{equation}
R \approx l_2 \sqrt{N_2}\label{e18}
\end{equation}
\begin{equation}
l_2 \approx l_3 \sqrt{N_3}\label{e19}
\end{equation}
\begin{equation}
R \sim l \sqrt{N}\label{e20}
\end{equation}
and a similar relation for the KBO (Kuiper Belt objects)
\begin{equation}
L \sim l_4 \sqrt{N_4}\label{e21}
\end{equation}
where $N_1 \sim 10^6$ is the number of superclusters in the
universe, $l_1 \sim 10^{25}cms$ is a typical supercluster size
$N_2 \sim 10^{11}$ is the number of galaxies in the universe and
$l_2 \sim 10^{23}cms$ is the typical size of a galaxy, $l_3 \sim
1$ light years is a typical distance between stars and $N_3 \sim
10^{11}$ is the number of stars in a galaxy, $R$ being the radius
of the universe $\sim 10^{28}cms, N \sim 10^{80}$ is the number of
elementary particles, typically pions in the universe and $l$ is
the pion Compton wavelength and $N_4 \sim 10^{10}, l_4 \sim 10^5
cm$, is the dimension of a typical KBO (with mass $10^{19}gm$ and $L$
the width of the Kuiper Belt $\sim 10^{10}cm$
cf.ref.\cite{cu}).\\
The size of the universe, the size of a supercluster etc. from
equations like (\ref{e17})-(\ref{e21}), as described in the
references turn up as the analogues of the Compton wavelength. For
example we have
\begin{equation}
R = \frac{h_1}{Mc}\label{e22}
\end{equation}
One can see that equations (\ref{e13}) to (\ref{e22}) are a consequence of gravitational orbits (or the Virial Theorem) and the conservation of angular momentum viz.,
\begin{equation}
\frac{GM}{L} \sim v^2, M v L = H\label{e23}
\end{equation}
(Cf.refs.\cite{BGSSU1,BGSSU2}), where $L,M,v$ represent typical length (or dispersion in length), mass and velocities at that scale and $H$ denotes the scaled Planck constant.\\
It also appears that equations (\ref{e17}) to (\ref{e21}) resemble a typical Random Walk relation (Cf.\cite{Rief}) of Brownian motion.\\
All this is suggestive but empirical. The question arises whether there is any theoretical justification. To investigate this further we observe that if we use (\ref{e23}) along with the relation,
$$L = vT$$
where $T$ is a typical time scale, for example the time period for an orbit, we get the relations
\begin{equation}
L^2 = \frac{H}{M} T \quad \left(H = \frac{GM^2}{v}\right)\label{e24}
\end{equation}
(\ref{e24}) is nothing but the well known equation of Nelson viz.,
\begin{equation}
\Delta x^2 = \nu \Delta t, \quad \nu = \frac{h}{m}\label{e25}
\end{equation}
where $\nu$ is the diffusion constant, $h$ the Planck constant and
$m$ the mass of a typical particle.\\
We now observe that as is well known,  the relations (\ref{e24}) or (\ref{e25}) lead to an equation identical to the Quantum Mechanical Schrodinger equation (Cf.ref.\cite{Nottale1} for a detailed derivation)
\begin{equation}
h_\imath \frac{\partial \psi}{\partial t} + \frac{h^2_\imath}{2m}
\nabla^2 \psi = 0\label{e26}
\end{equation}
Indeed this is not surprising because one can rewrite equation (\ref{e25}) as
\begin{equation}
m \Delta x \frac{\Delta x}{\Delta t} = h = \Delta x \cdot \Delta p\label{e27}
\end{equation}
which is the well known Uncertainty relation. Conversely, from the Uncertainty principle we could get back (\ref{e24}) or (\ref{e25}).\\
Interestingly it has been shown that this is true, not just for the special form of the diffusion constant, but also for any other form of the diffusion constant \cite{Davidson}. Another interesting point is that starting from (\ref{e24}) or (\ref{e13}), we can deduce equations like (\ref{e17}), which describe a Brownian path \cite{BGSPSP}.\\
In any case the steps leading to equation (\ref{e26}) and (\ref{e26}) itself provide the rationale for the scaled de Broglie or Compton lengths, for example equation (\ref{e22}), which follow from (\ref{e27}).\\
All this can be linked to Critical Point Theory and the Renormalization Group.For this, we observe that the creation of particles from a Quantum
vacuum (or pre space time) has been described in the references
cited \cite{r5a,r6}. It can be done within the context of the
above Nelsonian Theory, in complete analogy with the creation of
Benard cells at the critical point. In this development the Nelsonian-Brownian process
as described in (\ref{e25}) defines, first the Planck length, the
shortest possible length and then the random process leads to the
Compton scale (Cf.ref.\cite{BGSPSP}). This process is as noted
(Cf.ref.\cite{r6}) a complete analogue of the phase transition
associated with the Landau-Ginsburg equation \cite{r7}
\begin{equation}
-\frac{h^2}{2m} \nabla^2 \psi + \beta |\psi |^2 \psi = -\propto
\psi\label{e28}
\end{equation}
The parallel is not yet fully apparent, if we compare (\ref{e28})
and the Schrodinger equation (\ref{e25}). However this becomes
clear if we consider how the Schrodinger equation itself can be
deduced from the amplitudes of the Quantum vacuum, in which case
we get
\begin{equation}
\imath \hbar \frac{\partial \psi}{\partial t} =
\frac{-\hbar^2}{2m'}\frac{\partial^2 \psi}{\partial x^2} + \int
\psi^* (x')\psi (x)\psi (x')U(x')dx',\label{e29}
\end{equation}
Infact the correlation length from (\ref{e28}) is given by
$$\xi = (\frac{\gamma}{\propto})^{\frac{1}{2}} (\gamma \equiv \hbar^2/2m)$$
which can be easily reduced to the Compton wavelength. In other
words, the Schrodinger equation (\ref{e26}), via (\ref{e29})
describes the creation of particles,
a la Benard cells in a Landau-Ginsburg like phase transition.\\
As is known, the interesting aspects of the critical point theory
(Cf.ref.\cite{r7}) are universality and scale. Broadly, this
means that diverse physical phenomena follow the same route at the
critical point, on the one hand, and on the other this can happen
at different scales, as exemplified for example, by the course
graining techniques of the Renormalization Group. To highlight
this point we note that in critical point phenomena we have the
reduced order parameter $\bar Q$ and the reduced correlation
length $\bar \xi$. Near the critical point we have relations
like
$$(\bar Q) = |t|^\beta , (\bar \xi) = |t|^{-\nu}$$
Whence
\begin{equation}
\bar Q^\nu = \bar \xi^\beta\label{e30}
\end{equation}
In (\ref{e30}) typically $\nu \approx 2\beta$. As $\sqrt{Q} \sim
\frac{1}{\sqrt{N}}$ because $\sqrt{N}$ particles are created
fluctuationally, given $N$ particles, and in view of the fractal
two dimensionality of the path
$$\bar Q \sim \frac{1}{\sqrt{N}}, \bar \xi = (l/R)^2$$
This gives
$$R = \sqrt{N}l$$
which is nothing but (\ref{e20}).\\
In other words the scaled Planck effects and the scaled Random
Walk effects as typified by equations like (\ref{e13})-(\ref{e21})
are the result of a critical point phase transition and subsequent
course graining.\\
1. We observe that a Schrodinger equation like procedure has been used though in an empirical way by Agnese and Festa \cite{Agnese} to derive a Titius-Bode type relation for planetary distances which now appear as quantized levels. This consideration has been extended in an empirical way to also account for quantized cosmic distances \cite{Carneiro}.\\
Interestingly if we consider a wave packet of the generalized Schrodinger equation (\ref{e26}) with $h_1$ given by (\ref{e13}) for the universe itself, we have for a Gaussian wave packet
\begin{equation}
R \approx \frac{\sigma}{\sqrt{2}} \left(1 + \frac{h_1^2T^2}{\sigma^4 M^2}\right)^{1/2} \left(\approx \frac{1}{\sqrt{2}} \frac{h_1 T}{\sigma M}\right)\label{e31}
\end{equation}
where $R$ and $T$ denote the radius and age of the universe, $M$ its mass and $\sigma \sim R$ is the spread of the wave packet. As $R \approx cT$ (\ref{e31}) gives us back (\ref{e22}), that is the ``Compton wavelength'' of the universe treated as a wave packet.\\
2. Interestingly we can pursue the reasoning of
equations like  (\ref{e13}) to the case of terrestrial phenomena. Let us
consider a gas at standard temperature and pressure. In this case,
the number of molecules $n \sim 10^{23}$ per cubic centimeter, so
that $r \sim 1 cm$ and with the same $l$, we can get a "scaled"
Planck
constant $\tilde h \sim 10^{-44} << h$, the Planck constant.\\
In this case, a simple application of the WKB approximation, leads
immediately from the Schrodinger equation at the new scale to the
classical Hamilton-Jacobi theory, that is to classical mechanics.\\
3. Equations like (\ref{e17}) are the analogue of the well known Eddington formula
$$R = \sqrt{N}l$$
Similarly we can have the analogue of the mysterious Weinberg relation linking the pion mass to the Hubble constant, from $H^2 = M^3 LG$. For this we need to define the analogue of the Hubble constant $H$
$$\hat H = \frac{v}{L}$$
to get
$$M = \left(\frac{\hat H H^2}{Gv}\right)^{\frac{1}{3}}$$
which is the required relation.\\
4. We have argued that just as matter in the form of elementary particles, forms or condenses within the Compton wavelength from a background Quantum vaccuum in a phase transition, matter at other scales, for example stars and galaxies also condenses or clusters by a similar mechanism. This would give a rationale for the observed lumpyness of the universe. Similar considerations apply for the other scales referred to.


\begin{thebibliography}{99}
\bibitem {snyder1} H.S. Snyder, Physical Review, Vol.72, No.1, July 1 1947, p.68-71.
\bibitem {sny} H.S. Snyder, Physical Review, Vol.71, No.1, January 1 1947, p.38-41.
\bibitem {Wilson} K.G. Wilson, Rev.Mod.Phys., 1983, \underline{55}, pp.583ff.
\bibitem {kady} V.G. Kadyshevskii, Translated from Doklady Akademii Nauk SSSR, Vol.147, No.6 December 1962, p.1336-1339.
\bibitem {wolf} C. Wolf, Hadronic Journal, Vol.13, 1990, p.208-210.
\bibitem {fink} D.R. Finkelstein, ``Quantum Relativity A Synthesis of the Ideas of Einstein and Heisenberg'', Springer, Berlin, 1996.
\bibitem {cu} B.G. Sidharth, ``The Chaotic Universe: From the Planck to the Hubble Scale'', Nova Science Publishers, Inc., New York, 2001.
\bibitem {tooft} G.'t Hooft, Chemical and Quantum Gravity, \underline{13}, 1996,1023-1039.
\bibitem {mont}  I. Montway and G. Miinster, ``Quantum Fields on a Lattice'', Cambridge University Press, Cambridge, 1994, pp.164ff.
\bibitem {mup} B.G. Sidharth, Modified Uncertainty Principle
\bibitem {powell}J.L. Powell and B. Craseman, ``Quantum Mechanics'', Addision-Wesley, Cambridge, 1961.
\bibitem {mestres} L. Gonzales Mestres, Physics/9704017.
\bibitem {glashow} S. Coleman and S.L. Glashow, PRD, 59, 116008, 1999.
\bibitem {jacob} T. Jacobson, xxx.astro-ph/0212190.
\bibitem {olinto} A.V. Olinto, Phys.Rev., 333-334, 2000, pp.329ff.
\bibitem {carroll} S.M. Carroll, Phys.Rev.Lett., 87, 2001, pp.141601ff.
\bibitem {nag} M. Nagano, Rev.Mod.Phys., 72, 2000, pp.689ff.
\bibitem {Nottale1} L. Nottale, ``Fractal Space-Time and Microphysics: Towards a Theory of Scale Relativity'', World Scientific, Singapore, 1993, p.312.
\bibitem {Davis} H. Georgi in "The New Physics", Ed.P. Davies, Cambridge University Press, Cambridge, 1989, pp.446ff.
\bibitem {Nottale2} L. Nottale, Chaos, Solitons and Fractals, 1994, 4,3, p.361-388.
\bibitem {BGSSU1}B.G. Sidharth, Chaos, Solitons and Fractals, 12, 2001, 613-616.
\bibitem {BGSSU2} B.G. Sidharth, Chaos, Solitons and Fractals, 12, 2001, 1371-1373.
\bibitem {Rief} F. Reif, ``Statistical and Thermal Physics'', McGraw-Hill, Singapore, 1965.
\bibitem {Davidson} M. Davidson, Physica, 96A, p.465, 1979.
\bibitem {BGSPSP} B.G. Sidharth, Found.Phys.Lett., 15 (6), 2002, pp.577-583.
\bibitem {r5a} B.G. Sidharth, Chaos, Solitons and Fractals, 14 (2002), 1325-1330.
\bibitem {r6} B.G. Sidharth, Chaos, Solitons and Fractals, 18(1), September 2003, pp.197ff..
\bibitem {r7} D.L. Goodstein, "States of Matter", Dover Publications, Inc., New York, 1975, p.462ff.
\bibitem {Agnese} A.G. Agnese and R. Festa, Phys.Lett.A., 227, 1997, p.165-171.
\bibitem {Carneiro} S. Carneiro, Foundations of Physics Letters, Vol.II, No.1, 1998, p.95-102.
\end{thebibliography}
\end{document}